# NEUTRON-PROTON RADII IN N$\approx$Z NUCLEI

N.Auerbach

School of Physics and Astronomy, Tel Aviv University, Tel Aviv 69978, Israel

Abstract: A simple formula is derived that describes how the Coulomb interaction affects the proton radius in nuclei. It determines the difference between neutron and proton radii in nuclei with N$\approx$Z. It also provides an estimate for the difference between the radii of the Z core neutrons and the protons in nuclei with a large neutron excess. The results obtained from the derived formula are compared with radii calculated in a Skyrme Hartree-Fock calculation.

There is an ongoing quest to determine the radius of the neutron distribution in nuclei. The recent advances in nuclear theory and some new experiments have given this field an additional impetus. The experimental studies are plagued by the model dependence when relating the observables to the values of neutron radii. A promising attempt to measure the neutron radius in $Pb^{208}$ is the parity-violating electron scattering experiment at JLab [1]. The many theoretical studies involve the use of parameters not always well known and are hampered by the lack of reliable experimental data on neutron radii that would allow a calibration of some parameters. The various studies are often expressed in terms of the difference between neutron and proton radii. Of course knowing this difference enables one to determine the neutron radius because the proton radius is readily available experimentally.

In the present article we will study the difference of neutron-proton radii in nuclei with the number of neutrons N equal to the number of protons Z. In these nuclei the symmetry energy does not play any significant role, the only part that affects protons differently than neutrons are the charge asymmetric parts of the Hamiltonian, and of these by far the dominant one is the Coulomb interaction. It is of course expected that in such nuclei the protons will have a lager radius than the neutrons, forming a proton "skin". We will see that this effect is not negligible in nuclei with a charge Z. The number of stable nuclei with N=Z is small; however with the development of radioactive beams it will be possible to study unstable proton rich nuclei with larger Zs and small N-Z.

Consider a nucleus with N-neutrons and Z-protons. Let $H$ be the total Hamiltonian describing the nucleus:

$$H = H_0 + V \tag{1}$$

where $H_0$ is the part of the Hamiltonian that conserves isospin symmetry and $V$ contains all the parts of the total Hamiltonian that do not conserve this symmetry. The ground state wave function (w.f.) of $H$ is denoted by $|\tilde{0}\rangle$.

Let us treat now $V$ in perturbation theory. We can than write:

$$|\tilde{0}\rangle = |0\rangle + \sum_{n \neq 0} \varepsilon_n |n\rangle \qquad (2)$$

where $|0\rangle$ and $|n\rangle$ are the ground and excited states of the $H_0$ Hamiltonian.

The major part of the isospin violating interaction is the Coulomb force and of it the dominant one is the one-body Coulomb potential [2]. Let us simplify this potential by using a potential derived from a homogenous spherical charge distribution with radius $R$. Inside the sphere $(r \leq R)$:

$$V_C(r) = -\frac{Ze^2}{R^3} \sum_i \left( \frac{1}{2} r_i^2 - \frac{3}{2} R^2 \right) \left( \frac{1}{2} - t_z(i) \right) \qquad (3)$$

We need only to deal with the isovector part $V_C^{(1)}$ of this potential because the isoscalar part is assumed to be contained in $H_0$. The non-constant part that contributes to the non-diagonal matrix element:

$$V_C^{(1)} = \frac{Ze^2}{2R^3} \sum_i r_i^2 t_z(i) \qquad (4)$$

Therefore:

$$\varepsilon_n = \frac{\langle 0 | V_C^{(1)} | n \rangle}{E_n - E_0} \qquad (5)$$

We should comment that by choosing a nucleus with $N \neq Z$ in the case of the $H_0$ Hamiltonian means that the isospin symmetry is spontaneously broken. A symmetry potential that is a result of the difference in the two-body interaction between two nucleons in the T=0 and T=1 states:

$$V_{sym} = \frac{U}{A}(N - Z) t_z \qquad (6)$$

is isospin breaking even though its origin is an isospin conserving part of the nucleon-nucleon interaction. Therefore even in absence of real isospin breaking parts in $H_0$ one should expect some small differences in the wave functions of the core protons and neutrons. By "core" we mean the Z-neutrons occupying the orbits that the protons occupy. For N=Z nuclei this isospin breaking is zero for a nucleus described by $H_0$. When the excess of neutrons is small this holds to a good approximation.

We will now use the notion of the isovector giant monopole (IVGM) [2, 3] in order to calculate the mixing coefficient $\varepsilon_n$ in eq. (2). The z component of the isovector monopole operator is:

$$\hat{M}_0^{(1)} = \sum_i r_i^2 t_z \qquad (7)$$

For the use with off-diagonal matrix elements we can write

$$V_C^{(1)} = \frac{Ze^2}{2R^3} \hat{M}_0^{(1)} \qquad (8)$$

An "ideal" IVGM is:

$$\left|M_0^{(1)}\right\rangle = \frac{\sum_i r_i^2 t_z(i)|0\rangle}{\Omega} \equiv \frac{\hat{M}_0^{(1)}|0\rangle}{\Omega} \qquad (9)$$

where $\Omega$ is a normalization constant.

The state above exhausts the entire isovector monopole strength $r^2 t_z$. Clearly this state is not an eigenstate of the system. The strength is spread over several MeV, however it is still concentrated in a relatively narrow energy region. In our approach the above state is treated as a doorway. For more discussion see references [2, 4].

Let us now use the energy weighted sum rule (EWSR);

$$\frac{1}{2}\langle 0|[Q,[H,Q]]|0\rangle = \sum_n (E_n - E_0)\langle n|Q|0\rangle^2 \qquad (10)$$

Applying this to the $\hat{M}_0^{(1)}$ operator we get:

$$\frac{\hbar^2}{2m} A <r>^2 (1+\kappa) = \sum_n (E_n - E_0)\left|\langle n|\sum_i r_i^2 t_z(i)|0\rangle\right|^2 \qquad (11)$$

which for $V_C^{(1)}$ becomes:

$$(\frac{Ze^2}{2R^3})^2 \frac{\hbar^2}{2m} A <r^2> (1+\kappa) = \sum_n (E_n - E_0)\left|\langle 0|V_C^{(1)}(r_i)|n\rangle\right|^2 \qquad (12)$$

$\kappa$ in the above equations is the exchange correction, see [2,4,5].

We now assume that the sum can be exhausted by the single state $\left|M_0^{(1)}\right\rangle$, replacing the sum with a single term we can write:

$$(\frac{Ze^2}{2R^3})^2 \frac{\hbar^2}{2m} A <r^2> (1+\kappa) = (E_M - E_0)\left|\langle 0|V_C^{(1)}|M_0^{(1)}\rangle\right|^2 \qquad (13)$$

Let us now calculate the expectation value of $\hat{M}_0^{(1)}$ with the wave function $|\tilde{0}\rangle$ to first order in $\varepsilon_n$, using eq. (2)

$$\langle\tilde{0}|\sum_i r_i^2 t_z(i)|\tilde{0}\rangle \equiv \frac{1}{2}\left[N<\tilde{r}_n^2> - Z<\tilde{r}_p^2>\right] = \langle 0|\sum_i r_i^2 t_z(i)|0\rangle + 2\sum_{n\neq 0}\varepsilon_n\langle n|\sum_i r_i^2 t_z(i)|0\rangle \qquad (14)$$

Where $<\tilde{r}_n^2>$ and $<\tilde{r}_p^2>$ are respectively the neutron and proton mean square (m.s.) radii evaluated with the $|\tilde{0}\rangle$ wave function.

The right hand side (r.h.s.) of the above equation can be written as:

$$r.h.s. = \frac{1}{2}\left[N<r_n^2> - Z<r_p^2>\right] + \frac{Ze^2}{R^3}\sum_{n\neq 0}\frac{\left|\langle n|\hat{M}_0^{(1)}|0\rangle\right|^2}{E_0 - E_n} \qquad (15)$$

where $<r_n^2>$ and $<r_p^2>$ are the m.s. radii of neutrons and protons evaluated with $|0\rangle$.

We now will deal with the second term in the above equation. Using the EWSR and the doorway hypothesis we can write:

$$\frac{Ze^2}{R^3} \sum_{n \neq 0} \frac{|\langle n | \hat{M}_0^{(1)} | 0 \rangle|^2}{E_0 - E_n} = \frac{Ze^2}{R^3} \frac{\hbar^2}{2m} A \frac{<r^2>(1+\kappa)}{(E_0 - E_M)^2} \qquad (16)$$

We take now $R = 1.2 A^{1/3} fm$, for a homogenous charge distribution $<r^2> = \frac{3}{5} R$, we use $\kappa \approx 0.3$ as found in a number of calculations [4, 5] and take $E_M - E_0 \approx 140 A^{-1/3} MeV$ [5, 6].

With these reasonable choices and for nuclei with $N \approx Z$ we find for the *r.h.s.* of eq. (16) the simplified expression:

$$r.h.s. = 2. \times 10^{-3} Z^{7/3} \, fm^2 \qquad (17)$$

Therefore we can write:

$$\frac{1}{2}\left[N(<\tilde{r}_n^2> - <r_n^2>) - Z(<\tilde{r}_p^2> - <r_p^2>)\right] = -2. \times 10^{-3} Z^{7/3} fm^2 \qquad (18)$$

(the radii are expressed in fm).

We now write $N<\tilde{r}_n^2> = (N-Z)<\tilde{r}_{exc}^2> + Z<\tilde{r}_{n,c}^2>$ and $N<r_n^2> = (N-Z)<r_{ex}^2> + Z<r_{n,c}^2>$ where $<\tilde{r}_{ex}^2>$ denotes the m.s. radius of the excess neutrons and $<\tilde{r}_{n,c}^2>$ of the core neutrons (similarly for $<r_{ex}^2>$ and $<r_{n,c}^2>$) [4]. We can now write the left hand side *l.h.s.* of eq. (18):

$$l.h.s. = \frac{1}{2} Z \left[<\tilde{r}_{n,c}^2> - <r_p^2>\right] + \frac{N-Z}{2} \left[<\tilde{r}_{ex}^2> - <r_{ex}^2>\right] - \frac{Z}{2}\left[<r_{n,c}^2> - <r_p^2>\right] \qquad (19)$$

We will now droop the last term in this equation because in the absence of Coulomb mixing and for nuclei with no or a small neutron excess the two m.s. radii are equal. Denoting differences in the root mean square radii (r.m.s.):

$$\delta r_{np} = <\tilde{r}_{n,c}^2>^{1/2} - <\tilde{r}_p^2>^{1/2} \text{ and } \delta r_{ex} = <\tilde{r}_{ex}^2>^{1/2} - <r_{ex}^2>^{1/2} \qquad (20)$$

we find:

$$\delta r_{np} + \frac{(N-Z)}{Z} \frac{r_{ex}}{r_p} \delta r_{ex} = -1.6 \times 10^{-3} Z \, fm \qquad (21)$$

For $N = Z$ nuclei or nuclei with a small neutron excess this formula reduces to:

$$\delta r_{np} = -1.6 \times 10^{-3} Z \, fm \qquad (22)$$

The difference between the r.m.s radii of neutrons and protons in such nuclei is negative, as one should expect, the Coulomb force pushes the protons away. The dependence on the charge of the nucleus is simple and the difference increases with Z.

For $O^{16}$ the neutron r.m.s. radius is smaller than the proton one by 0.01 fm, in $Ca^{40}$ this difference is -0.03 fm, in $Ni^{56}$ it is -0.04 fm and in $Sn^{100}$ it becomes -0.08 fm. A Skyrme Hartree-Fock calculation for $Ca^{40}$ [4] results in $\delta r_{np} = -0.04$ $fm$.
In nuclei with a large neutron excess one can compute using a realistic Skyrme HF, the r.m.s. for the Z-core neutrons (that is to exclude the excess neutrons) and Z-protons and then find $\delta r_{np} = <\tilde{r}_{n,c}^2>^{1/2} - <\tilde{r}_p^2>^{1/2}$. The results are shown in table 1.

Table1. The values of $\delta r_{np} = <\tilde{r}_{n,c}^2>^{1/2} - <\tilde{r}_p^2>^{1/2}$ in fm.

| Nucleus | eq. (22) | HF |
| --- | --- | --- |
| $Ca^{40}$ | -0.03 | -0.04 |
| $Sr^{88}$ | -0.06 | -0.10 |
| $Ce^{140}$ | -0.09 | -0.11 |
| $Pb^{208}$ | -0.13 | -0.14 |

The results for nuclei with the large neutron excess are indicative of the effect Coulomb repulsion has on the difference between neutron and proton radii in the core. The neutron radii for the core nucleons are smaller than the proton radii. Of course the excess neutrons occupy higher orbits with larger and larger radii and the total neutron r.m.s. radius is larger that the proton r.m.s. radius. However it is clear that the Coulomb repulsion mitigates the difference in the neutron and proton radii. Equation (22) is valid to a very good approximation for N=Z nuclei, and to a good approximation for nuclei with a very small ratio $(N-Z)/Z$.
The proton skin in nuclei like $Sn^{100}$ is significant, the proton radius is about 1.5% larger then the neutron radius.
Comparison to experiment is difficult. As already mentioned there is little reliable data for neutron radii, especially for nuclei with a small neutron excess.
In an experiment performed with anti-protonic atoms [7], $\delta r_{np}$ is negative for $Ca^{40}$, $Ni^{58}$ and $Ni^{60}$. There error bars in this measurements are too large to make this comparison definite.

Even a small proton "skin" in N=Z nuclei can make a difference in certain processes that occur at the surface. For example, the interaction of medium energy (~180 MeV) pions with nuclei is an example of this. A paper published 28 years ago [8] discussed the results of the observation of the two isospin components $\Delta T_z = \pm 1$ of the isovector dipole in $Ca^{40}$ in the charge exchange reactions $(\pi^{\pm}, \pi^0)$ [9]. The energies of the pions were about 164 MeV, thus in the strongly absorbing regime. The cross section for the $(\pi^-, \pi^0)$ reaction was 70% larger than in the $(\pi^+, \pi^0)$ process. The transition strength $S_{\pm}$ to the two components of the isovector dipole satisfies the relation [5, 8]:

$$(S_- - S_+) \sim (N <r_n^2> - Z <r_p^2>) \qquad (23)$$

As we have seen in $Ca^{40}$ the proton radius is larger than the neutron radius by 0.03-0.04 fm. It was shown in [8] that this was enough to make a big difference in the pion cross-sections because of the surface nature of the two charge-exchange reactions.